\let\accentvec\vec
\documentclass{llncs}

\let\vec\accentvec
\usepackage{graphicx}
\usepackage[utf8]{inputenc}
\usepackage{epstopdf}
\usepackage{multirow,bigstrut}
\usepackage{amssymb,amsmath,bm}
\usepackage{subfigure}
\usepackage{tabu}
\usepackage{cite}
\usepackage{url}
\usepackage{cite}
\usepackage{algorithm}
\usepackage{algorithmic}
\usepackage{stmaryrd}
\usepackage{amsfonts}

\begin{document}

\title{Design and evaluation of chaotic iterations based keyed hash function}
\author{Zhuosheng Lin\inst{1}, Christophe Guyeux\inst{2}, Simin Yu\inst{1} \and Qianxue Wang\inst{1}}
\institute{College of Automation, Guangdong University of Technology, Guangzhou, China \\
    \email{zhuoshenglin@163.com, siminyu@163.com, wangqianxue@gdut.edu.cn}
	\and Femto-st Institute, University of Bourgogne Franche-Comt\'{e}, Besan\c{c}on, France \\
    \email{christophe.guyeux@univ-fcomte.fr}}

\maketitle

\begin{abstract}
	Investigating how to construct a secure hash algorithm needs in-depth study, as various existing hash functions like the MD5 algorithm have recently exposed their security flaws. At the same time, hash function based on chaotic theory has become an emerging research in the field of nonlinear information security. As an extension of our previous research works, a new chaotic iterations keyed hash function is proposed in this article. Chaotic iterations are used both to construct strategies with pseudorandom number generator and to calculate new hash values using classical hash functions. It is shown that, by doing so, it is possible to apply a kind of post-treatment on existing hash algorithms, which preserves their security properties while adding Devaney's chaos. Security performance analysis of such a post-treatment are finally provided.
\end{abstract}


\section{Introduction}
A hash function is any function that can be used to map data of arbitrary size to data of fixed size, which has many information-security applications. Rivest designed the first famous hash function called MD4 (Message Digest 4) in 1990, which is based on Merkle-Damgard iterative structure~\cite{Rivest1990The}. Later, various hash functions with an improved but similar design have been proposed. The latter encompass the well-known MD5~\cite{Rivest1992The} and SHAs secure hash algorithm series~\cite{US1995Secure}. However, recent researches have shown that security flaws exist too in these widely used standard hash functions. For instance, Lenstra, cooperated with Xiaoyun Wang, forged a digital certificate with different keys~\cite{Stevens2007Chosen}. Then they improved the MD5 collision course and constructed an effective certificate~\cite{Stevens2009Short}. This research result shocked cryptologists.
%MD5 and SHA-0 have been broken a long time ago \cite{wang2005efficient}. Furthermore, attack over SHA-1 has been achieved needing only $2^{69}$ operations \cite{wang2005finding}.

Some relationships can be emphasized between chaos properties and some targeted aims in cryptology. Thus it may be a good idea to investigate the use of chaos to enrich the design of cryptographic systems. In our previous work, we have proven that discrete chaotic iterations (CIs) produce topological chaos as described by Devaney~\cite{bahi2013discrete}. This topological chaos is a well studied framework and we have applied it in hash function, pseudorandom number generation, data hiding, and so on. However, all of them are used separately. In this research work, we intend to combine pseudorandom number generation and hash functions using CIs. Then we will check if this combination can improve the security performance of standard hash functions. More precisely, we will apply chaotic iterations on classical hash functions, adding by doing so provable chaos while preserving security properties like the collision. With such chaos, our desire is to reinforce diffusion and confusion of the inputted hash functions. In the meantime, General Formulation of the Chaotic Iterations (GFCIs) will be introduced and used, to deal with the output of standard hash functions and to construct chaotic strategies.

The remainder of this article is organized as follows. The first next section is devoted to some basic recalls on the general form of chaotic iterations. The third section introduces pseudorandom number generator with CIs. Our CI-based hash function is proposed and reformulated in this section too. Experimental evaluation is shown in the fourth section. This article ends by a conclusion section, in which our research work is summarized.

\section{General Formulation of the Chaotic Iterations}
In this section, we focus on the general formulation of chaotic iterations. Let us first define some notations.  $\mathbb{N}$ is the set of natural (non-negative) numbers. The domain $\mathbb{N}^*= \left\lbrace 1,2,3,\ldots \right\rbrace$ is the set of positive integers and $\mathbb{B} = \left\lbrace 0,1 \right\rbrace$. $\left[\kern-0.15em \left[ 1;N \right]\kern-0.15em \right]= \left\lbrace 1,2,3,\ldots,N \right\rbrace$. A sequence which elements belong in $\left[\kern-0.15em \left[ 1;N \right]\kern-0.15em \right]$ is called a strategy. The set of all strategies is denoted by $\mathit{S}$. $\mathit{S^n}$ denotes the $\mathit{n^{th}}$ term of a sequence $\mathit{S}$, $\mathit{X_i}$ stands for the $\mathit{i^{th}}$ components of a vector $\mathit{X}$.

In here a new kind of strategies is introduced, namely a sequence of subsets of $\llbracket 1, N \rrbracket$, that is, a sequence of $\mathcal{P}(\llbracket 1, N\rrbracket)^\mathbb{N}$, where $\mathcal{P}\left(X\right)$ is for the powerset of the set $X$ (\textit{i.e.}, $Y \in \mathcal{P}\left(X\right) \Longleftrightarrow Y \subset X$). So we can now change multiple bits between two adjacent outputs, as follows.

The general form of  the discrete dynamical system in chaotic iterations is
\begin{equation}
\begin{array}{c}
x^0 \in \mathbb{B}^N, (S^n)_{n \in \mathbb{N}} \in \mathcal{P}(\llbracket 1, N \rrbracket)^\mathbb{N}\\
\forall n \in \mathbb{N}^*, \forall i \in \left[\kern-0.15em \left[ 1;N \right]\kern-0.15em \right],x^n_i =
\begin{cases}
x^{n-1}_i, & \mbox{if } i \notin S^n \\
\left( f\left( x^{n-1}\right) \right)_{S^n}, & \mbox{if } i \in S^n
\end{cases}
\end{array}{}
\end{equation}

In other words, at the $n^{th}$ iteration, only the cells whose id is contained into the set $S^{n}$ are iterated.

Let us now rewrite these general chaotic iterations as usual discrete dynamical system of the form $X^{n+1}=f(X^n)$ on an \textit{ad hoc} metric space. Such a formulation is required in order to study the topological behavior of the system.

Let us introduce the following function:
\begin{equation}
\begin{array}{c}
\psi: \llbracket 1;N \rrbracket  \times \mathcal{P}\left(\left[\kern-0.15em \left[ 1;N \right]\kern-0.15em \right]\right)  \longrightarrow  \mathbb{B}\\
\left(i,X\right) \longleftarrow
\begin{cases}
0 & \mbox{if } i \notin X, \\
1 & \mbox{if } i \in X.
\end{cases}
\end{array}{}
\end{equation}

Given a function $f:\mathbb{B}^N \longrightarrow \mathbb{B}^N $, define the function:
\begin{equation}
\begin{array}{c}
F_{f}: \mathcal{P}\left(\left[\kern-0.15em \left[ 1;N \right]\kern-0.15em \right] \right) \times \mathbb{B}^{N}
\longrightarrow  \mathbb{B}^{N}, \\
(P,E)  \longmapsto \left( E_{j} \cdot \psi (j,P)+f(E)_{j} \cdot \overline{\psi
	(j,P)}\right) _{j\in \left[\kern-0.15em \left[ 1;N \right]\kern-0.15em \right]}
\end{array}.
\end{equation}
Consider the phase space:
\begin{equation}
\mathcal{X} = \mathcal{P}\left(\left[\kern-0.15em \left[ 1;N \right]\kern-0.15em \right]\right)^\mathbb{N} \times
\mathbb{B}^N,
\end{equation}
\noindent and the map defined on $\mathcal{X}$:
\begin{equation}
G_f\left(S,E\right) = \left(\sigma(S), F_f(i(S),E)\right), \label{Gf}
\end{equation}
where, in a similar formulation than previously, $\sigma$ is the \emph{shift} function defined by $\sigma:
(S^{n})_{n\in \mathbb{N}}\in \mathcal{P}\left(\left[\kern-0.15em \left[ 1;N \right]\kern-0.15em \right]\right)^\mathbb{N}\longrightarrow (S^{n+1})_{n\in \mathbb{N}}\in \mathcal{P}\left(\left[\kern-0.15em \left[ 1;N \right]\kern-0.15em \right]\right)^\mathbb{N}$ and $i$ is the \emph{initial function}
$i:(S^{n})_{n\in \mathbb{N}} \in \mathcal{P}\left(\left[\kern-0.15em \left[ 1;N \right]\kern-0.15em \right]\right)^\mathbb{N}\longrightarrow S^{0}\in \mathcal{P}\left(\left[\kern-0.15em \left[ 1;N \right]\kern-0.15em \right]\right)$.
Then the general chaotic iterations defined in Equ.6 can
be described by the following discrete dynamical system:
\begin{equation}
\left\{
\begin{array}{l}
X^0 \in \mathcal{X} \\
X^{k+1}=G_{f}(X^k).%
\end{array}%
\right.
\end{equation}%

To study the Devaney's chaos property, a relevant distance between two points
$X = (S,E), Y = (\check{S},\check{E})$ of $\mathcal{X}$ must be defined.
Let us introduce:
\begin{equation}
d(X,Y)=d_{e}(E,\check{E})+d_{s}(S,\check{S}),
\end{equation}
where
\begin{equation}
\left\{
\begin{array}{lll}
\displaystyle{d_{e}(E,\check{E})} & = & \displaystyle{\sum_{k=1}^{\mathsf{N}%
	}\delta (E_{k},\check{E}_{k})}\textrm{ is once again the Hamming distance}, \\
\displaystyle{d_{s}(S,\check{S})} & = & \displaystyle{\dfrac{9}{\mathsf{N}}%
	\sum_{k=1}^{\infty }\dfrac{|S^k\Delta {S}^k|}{10^{k}}}.%
\end{array}
\right.
\end{equation}
where $|X|$ is the cardinality of a set $X$ and $A\Delta B$ is for the symmetric difference, defined for sets A, B as $A\,\Delta\,B = (A \setminus B) \cup (B \setminus A)$.

It has been proven in \cite{Guyeux2015Efficient} that:
\begin{theorem}
	The general chaotic iterations defined on Equ.1 satisfy the Devaney's property of chaos.
\end{theorem}

\section{Security tools based on CIs}
We now investigate how to apply chaotic iterations on existing security tools. By such kind of post-treatment, we will add chaos to these tools, hoping by doing so to improve them in practice (increasing the entropy of random generators, the diffusion and confusion of hash functions, etc.) Such improvement must be such that existing security properties are preserved through iterations.
\subsection{Pseudorandom number generator with CIs}
In this section, we consider that the strategy $(S^n)_{n \in \mathbb{N}}$ is provided by a pseudorandom number generator, leading to a collection of so-called CIPRNGs~\cite{guyeux15a:it}. The XOR CIPRNGs, for instance, is defined as follows~\cite{Guyeux2015Efficient}:
\begin{equation}
\begin{cases}
x^0 \in \mathbb{B}^N\\
\forall n\in \mathbb{N}^*, x^{n+1}=x^n \oplus S^n,
\end{cases}
\end{equation}
where $N\in\mathbb{N}^*$ and $\oplus$ stands for the bitwise exclusive or (xor) operation between the binary decomposition of $x^n$ and $S^n$.
In the formulation above, chaotic strategy $(S^n)_{n\in\mathbb{N}^*} \in \left[\kern-0.15em \left[ 1;N \right]\kern-0.15em \right]^N$ is a sequence produced by any standard pseudorandom number generator, which can be the well-known Blum Blum Shub (B.B.S.), Linear Congruential Generator (LCG), Mersenne Twister (MT), XORshift, RC4, or the Linear-Feedback Shift Register (LFSR). %Lastly, iterate operation is bitwise exclusive xor operation.
XOR CIPRNGs, which can be written as general chaotic iterations using the vectorial negation (see~\cite{Guyeux2015Efficient}), have been proven chaotic. They are able to pass all the most stringent statistical batteries of test, for well-chosen inputted generators.

%The proposed pseudorandom number generator based on chaotic iterations is designed by the following process. Initial state $x^0\in  \mathbb{B}^N$ is a $N$ bit Boolean vector chosen by the use.

\subsection{CIs-based hash function}
Let us now present our hash function $H_h:K\times\mathbb{B}^*\rightarrow\mathbb{B}^N$ that is based on GFCIs. The key $k=\{k_1,k_2,k_3\}$ is in key space $K$, where $k_1, k_2$, and $k_3$ are parameters of the function. The proposed hash function $H_h$ is realized as follows.

The first step of the algorithm is to choose the traditional hash function $h$ that it will be used in the proposed hash function. For our implementations, we have chosen MD5, SHA-1, SHA-224, SHA-256, SHA-384, and SHA-512. The selected hash function determines the length $N$ of the output hash value.

Then, the input message $x$ is needed to transform  into a normalized bits sequence of length a multiple of $N$, by applying the SHA-1 normalization stage. After this initialization, the length of the treated sequence $X$ is $L$.

In the third step, $k_1$ is used as seed to generate $k_2$ pseudorandom binary vectors of length $N$, with XOR CIPRNGs. %The generated pseudorandom numbers $m$ are used to construct the strategies $S$.
This sequence is the chaotic strategy %We split this binary flow sequence to be
$S=\left\lbrace S^0S^1S^2\ldots S^{k_2-1}\right\rbrace$. %, where each $S^i$ is a $N$ bit Boolean vector.

In the forth step, $k_3$ is considered as a binary vector of length $N$. $S^i\in S$ is then combined with $k_3$ using exclusive-or operation. After that we get a $N$ bit binary output. Then we split $X$ into $X=\left\lbrace X^0X^1\ldots\right\rbrace$. Each $X^i$ will be combined with the output of $S$ and $K_3$ using exclusive-or operation. After that, we use this result as the input of traditional hash function $h$.

Lastly, to construct the digest, chaotic iteration of $G_f$ are realized with the traditional hash function output $h(k_3,m,X)$ and strategy $S$ as defined above. The result of these iterations is a $N$ bits vector. It is translated into hexadecimal numbers to finally obtain the hash value.

So, the keyed hash function $H_h:K\times\mathbb{B}^{*}\rightarrow\mathbb{B}^{N}$ is described as Algorithm 1.
\begin{algorithm}
	\caption{ The proposed hash function $H_h$}
	\label{hash function}
	\begin{algorithmic}[1]
		\REQUIRE ~~\\
		The key $k=\left( k_1,k_2,k_3\right)\in K $;\\
		The input message $x\in\mathbb{B}^{*}$;\\
		The standard hash function $h()$;\\
		\ENSURE ~~\\
		Hash value $H_h$;
		\STATE Transform $x$ to sequence $X$ which length is $L$;
		\STATE Use XOR CIPRNGs to generate $m$ using $k_1$ as a seed and construct strategy $S=\left\lbrace S^0S^1\ldots S^{k_2-1}\right\rbrace$ with $m$
		\STATE Use standard hash function to generate hash value $H=h(k_3,m,X)$;
		\FOR{$i=0,1,2,\ldots,k_2-1$}
		\STATE Use GFCIs to generate hash value: $H_h=G_f(S^i,H)$;
		\ENDFOR
		\RETURN $H_h$;
	\end{algorithmic}
\end{algorithm}

$H_h$ is thus a chaotic iteration based post-treatment on the inputted hash function. If $h$ satisfies the collision resistance property, then it is the case too for $H_h$. Moreover, if $h$ satisfies the second-preimage resistance property, then it is the case too for $H_h$, as proven in \cite{Guyeux2014Introducing}. With this post-treatment, we can thus preserve security while adding chaos: the latter may be useful to improve both confusion and diffusion.

\section{Experimental Evaluation}

In this section, experimental evaluations are provided including hash values, diffusion and confusion, and crash analysis. Let us consider that the input message is the poem ``Ulalume" (E.A.Poe) and the selected pseudorandom number generator is B.B.S.

\subsection{Hash values}
 The standard hash function that we use here is MD5. To give illustration of the key properties, we will use the proposed hash function to generate hash values in the following cases:
\begin{itemize}
	\item Case 1. $k_1=50,k_2=2,k_3=50$, and B.B.S.
	\item Case 2. $k_1=51,k_2=2,k_3=50$, and B.B.S.
	\item Case 3. $k_1=50,k_2=3,k_3=50$, and B.B.S.
	\item Case 4. $k_1=50,k_2=2,k_3=51$, and B.B.S.
\end{itemize}

The corresponding hash values in hexadecimal format are:
\begin{itemize}
	\item Case 1. F69C3F042ABA1139FF443C278FDF3F7F.
	\item Case 2. C31BFBDD43273913C7CC845EC5E3D1EE.
	\item Case 3. 43353FA45B9560413C059F7FD4F485FB.
	\item Case 4. BEA4CAD480333117292F421BFA401BEB.
\end{itemize}

From simulation results, we can see that any little change in key space $K$ can cause a substantial modification in the final hash value, which is coherent with the topological properties of chaos. In other words, it seems to be extremely sensitive to initial parameters.

A secured hash function should not only be sensitive to initial parameters, but also to initial values. This is why we test now our hash function with some changes in the input message, and observe the distribution of hash values. The key we use here is $k_1=50,k_2=2,k_3=50$, and standard hash function is MD5.
\begin{itemize}
	\item Case 1. The input message is the poem ``Ulalume" (E.A.Poe).
	\item Case 2. We replace the last point `.' with a coma `,'.
	\item Case 3. In ``The skies they were ashen and sober", `The' become `the'.
	\item Case 4. In ``The skies they were ashen and sober", `The' become `Th'.
	\item Case 5. We add a space at the end of the poem.
\end{itemize}

The corresponding hash values in binary format are shown in Figure 1. Through this experiment, we can check that the proposed hash function is sensitive to any alteration in the input message, which will cause the modification of the hash value.
\begin{figure}[htb]
	\centering
	\includegraphics[width= 0.7\textwidth]{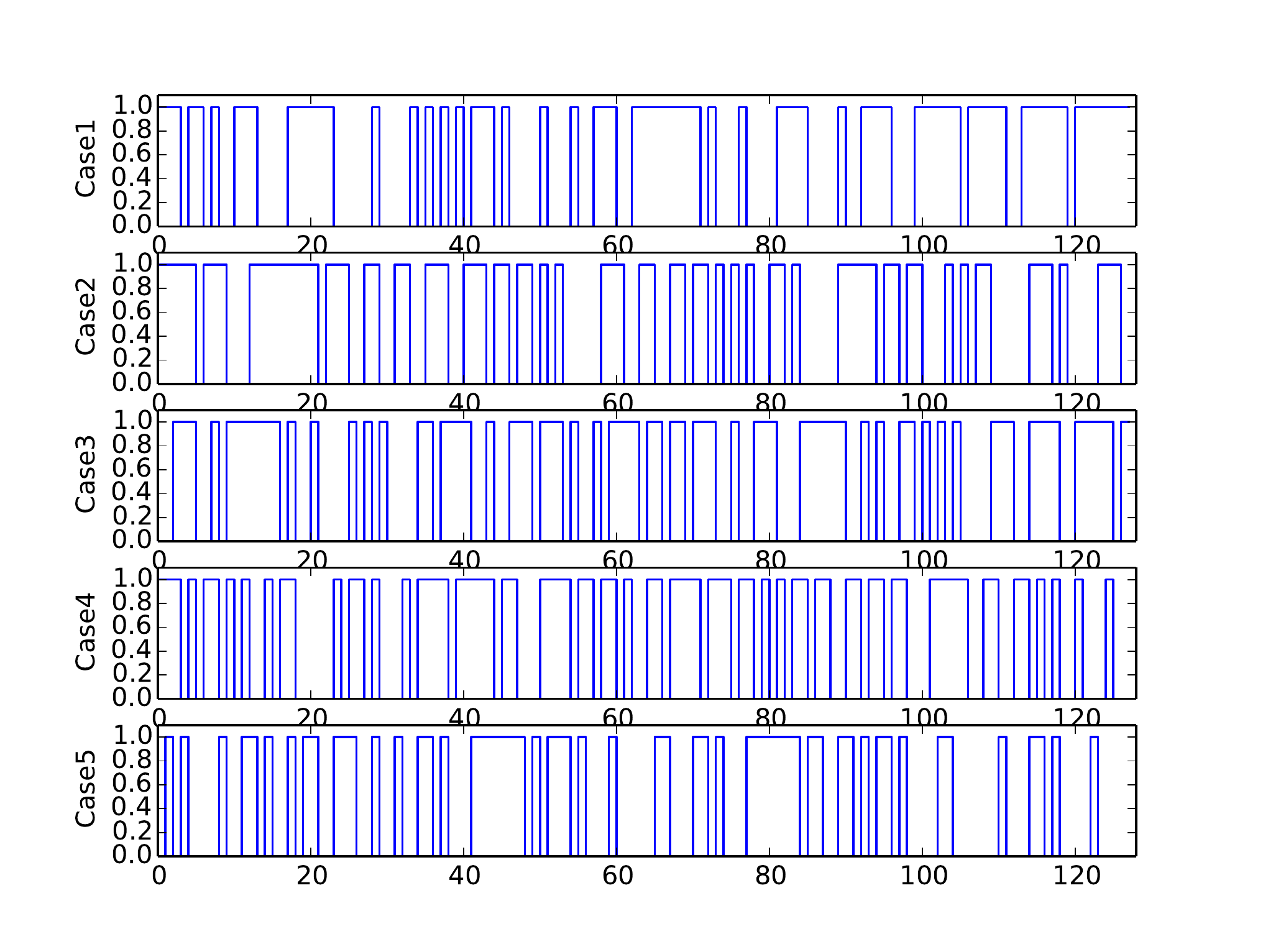}
	\caption{ 128 bit hash values in various cases }
	\label{fig1}
\end{figure}

 For a secured hash function, the repartition of its hash values should be uniform. In other words, the algorithm should make full use of cryptogram space to make that the hash values are evenly distributed across the cryptogram space. The cases here are the same as discussed above. In Figure 2(a), the ASCII codes of input message are localized within a small area, whereas in Figure 2(b), the hexadecimal numbers of the hash value are uniformly distributed in the area of cryptogram space.
\begin{figure}[htb]
	\centering
	\subfigure[][\emph{Original text (ASCII)}.]
	{\includegraphics[width=.35\textwidth]{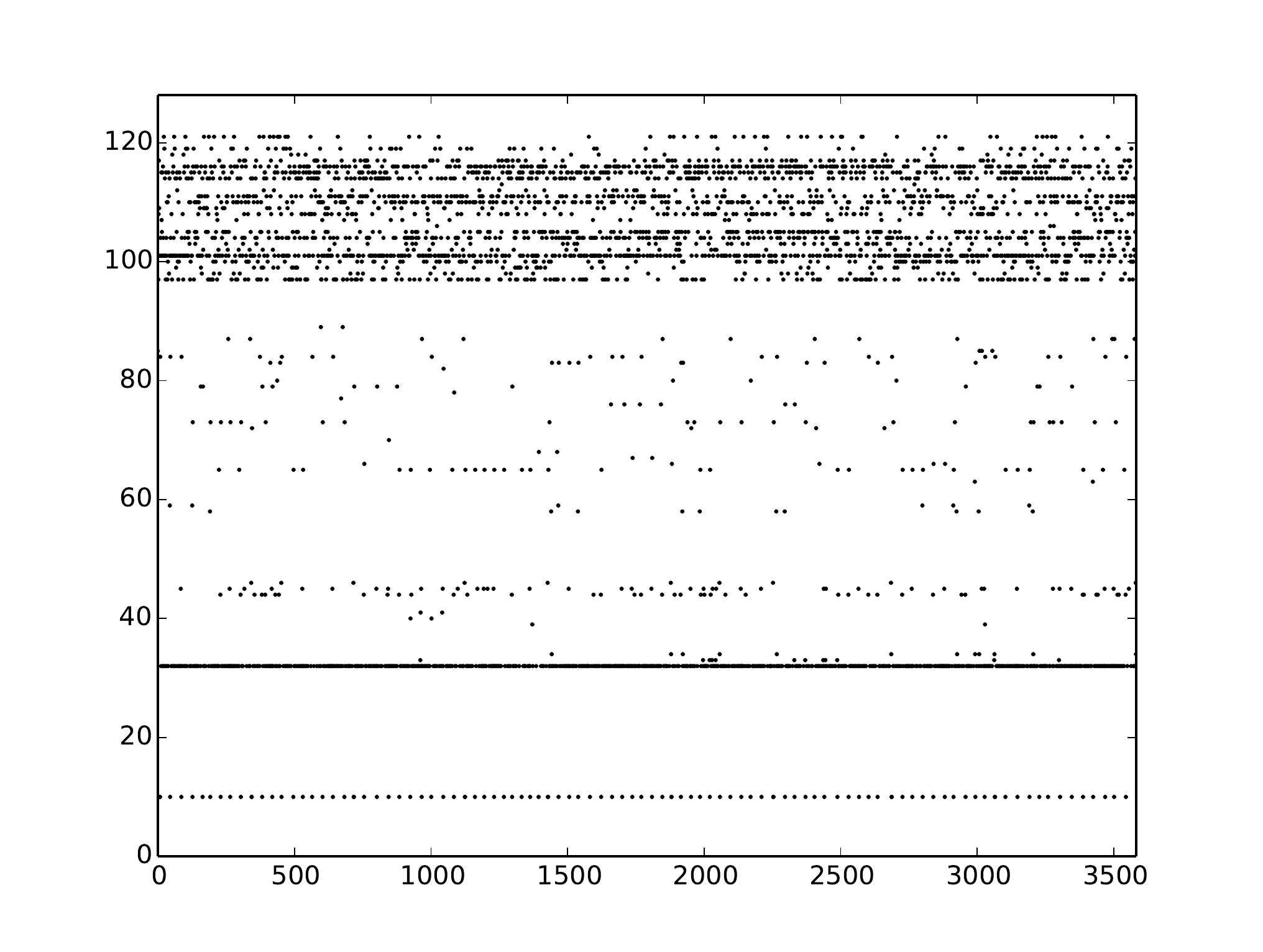}} \quad
	\subfigure[][\emph{Hash value use MD5}.]
	{\includegraphics[width=.35\textwidth]{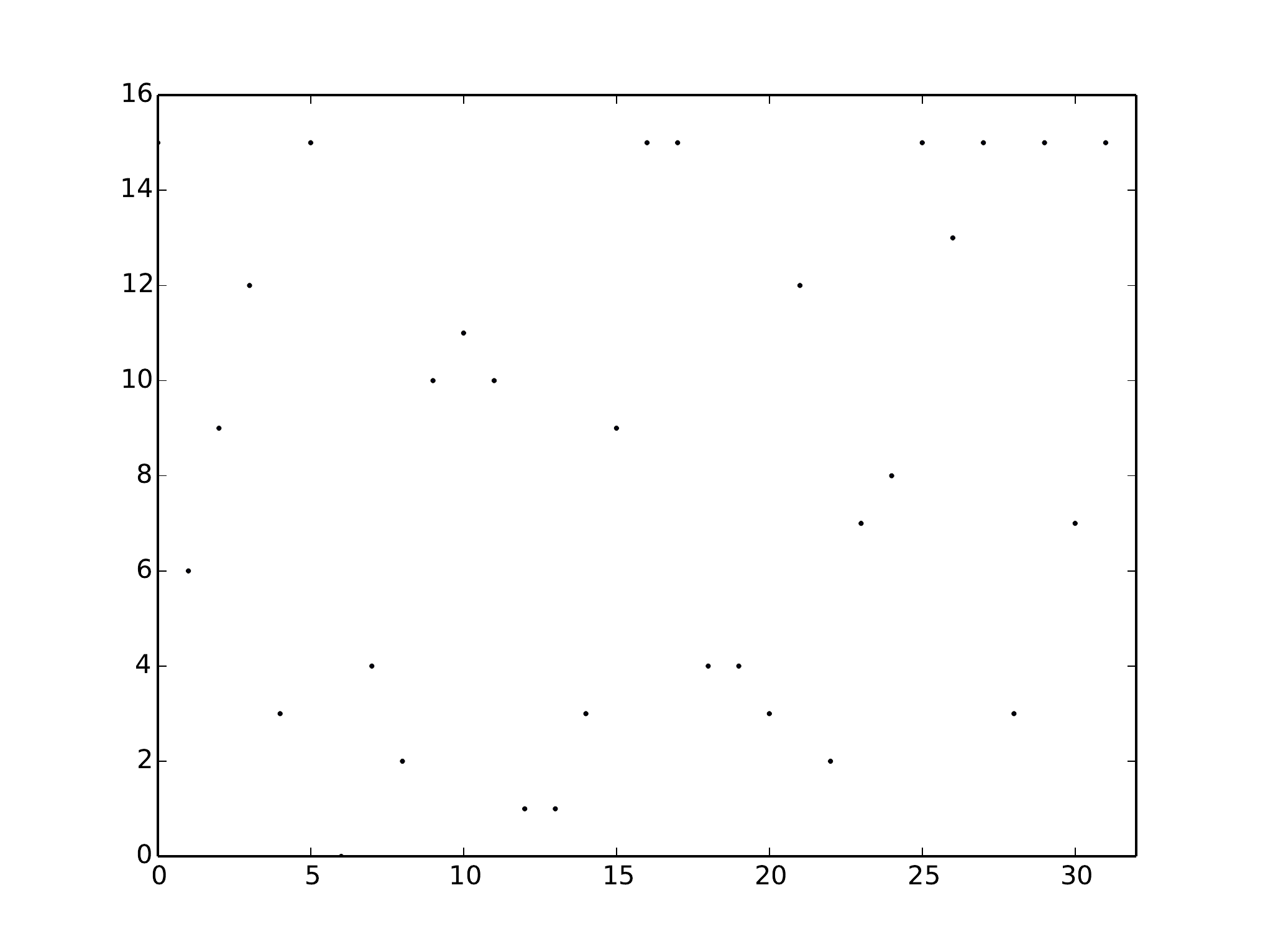}} \\
	\caption{Spread of input message and the corresponding hash value}
	\label{fig2}
\end{figure}

\subsection{Diffusion and Confusion}

We now focus on the illustration of the diffusion and confusion properties. Let us recall that diffusion means that the redundancy of the plain text must be dispersed into the space of cryptograms so as to hide the statistics of plain text. Confusion refers to the desire to make the statistical relationship between plain text, ciphertext, and keys as complex as possible, which makes attackers difficult to get relation about keys from ciphertext. So under the situation of that when the plain text is changed by only one bit, it leads to a modification of hash values that can be described by the following  statistics:
\begin{itemize}
	\item Mean changed bit number: $\overline{B}=\tfrac{1}{N}\sum_{i=1}^N B_i$;
	\item Mean changed probability: $P=\tfrac{\overline{B}}{L}\times100\%$;
	\item Mean square error of B: $\Delta B=\sqrt{\tfrac{1}{N-1}\sum_{i=1}^N(B_i-\overline{B})}$;
	\item Mean square error of P: $\Delta P=\sqrt{\tfrac{1}{N-1}\sum_{i=1}^N(\tfrac{B_i}{L}- P)}\times100\%$;
\end{itemize}
where $N$ denotes the statistical times, and $B_i$ denotes the changed bits of hash value in $i^{th}$ test, while $L$ denotes the bits of hash value in binary format. For a secured hash function, the desired value of $\overline{B}$ should be $L/2$. The desired distribution of hash algorithm should be that small toggle in plain text cause $50\%$ change of hash value. $\Delta B$ and $\Delta P$ show the stability of diffusion and confusion properties. The hash algorithm is more stable if these two values are close to $0$.

\begin{figure}
	\centering
	\subfigure[][\emph{Standard hash function is MD5}.]
	{\includegraphics[width=.45\textwidth]{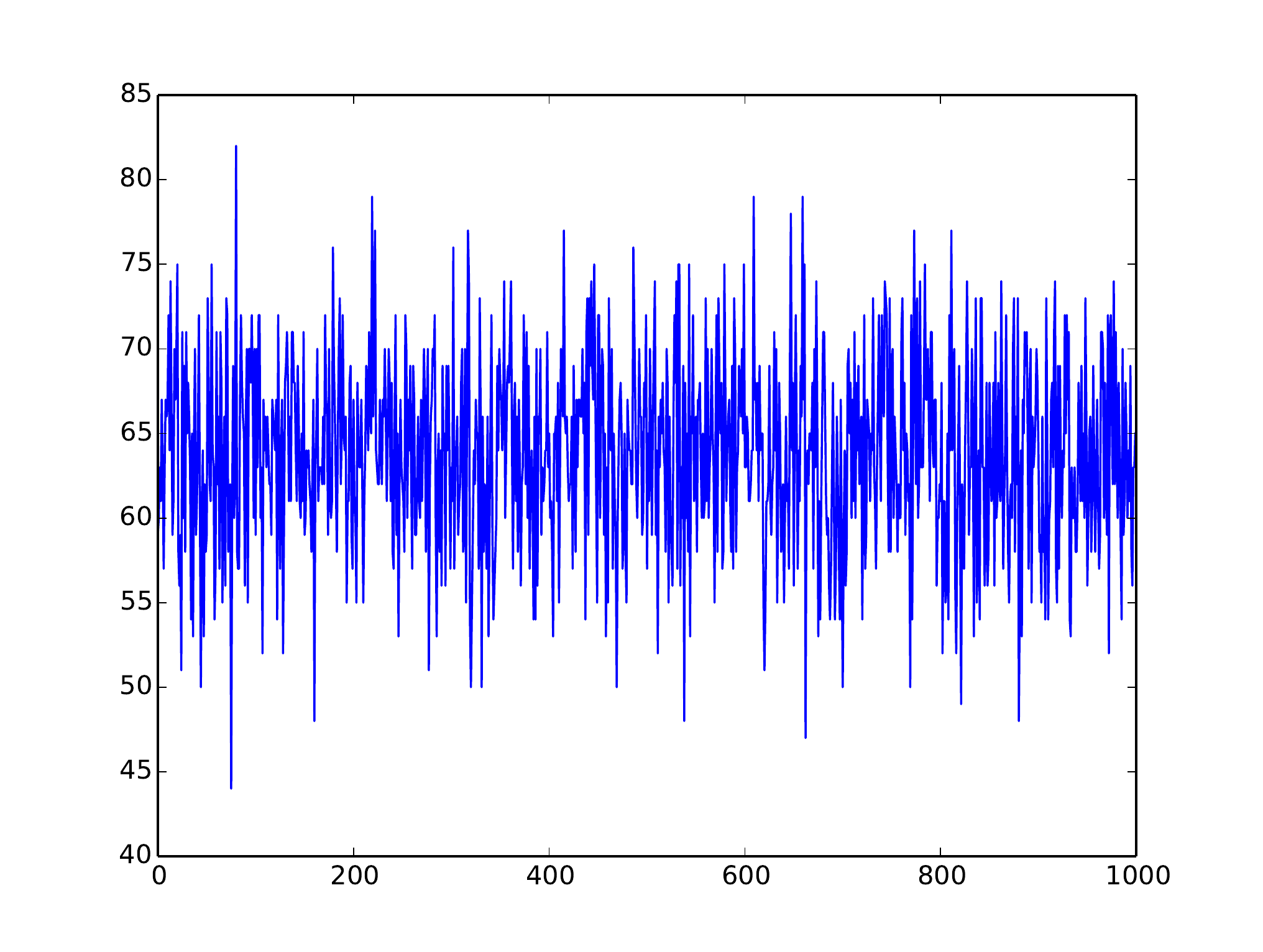}} \quad
	\subfigure[][\emph{Standard hash function is SHA-512}.]
	{\includegraphics[width=.45\textwidth]{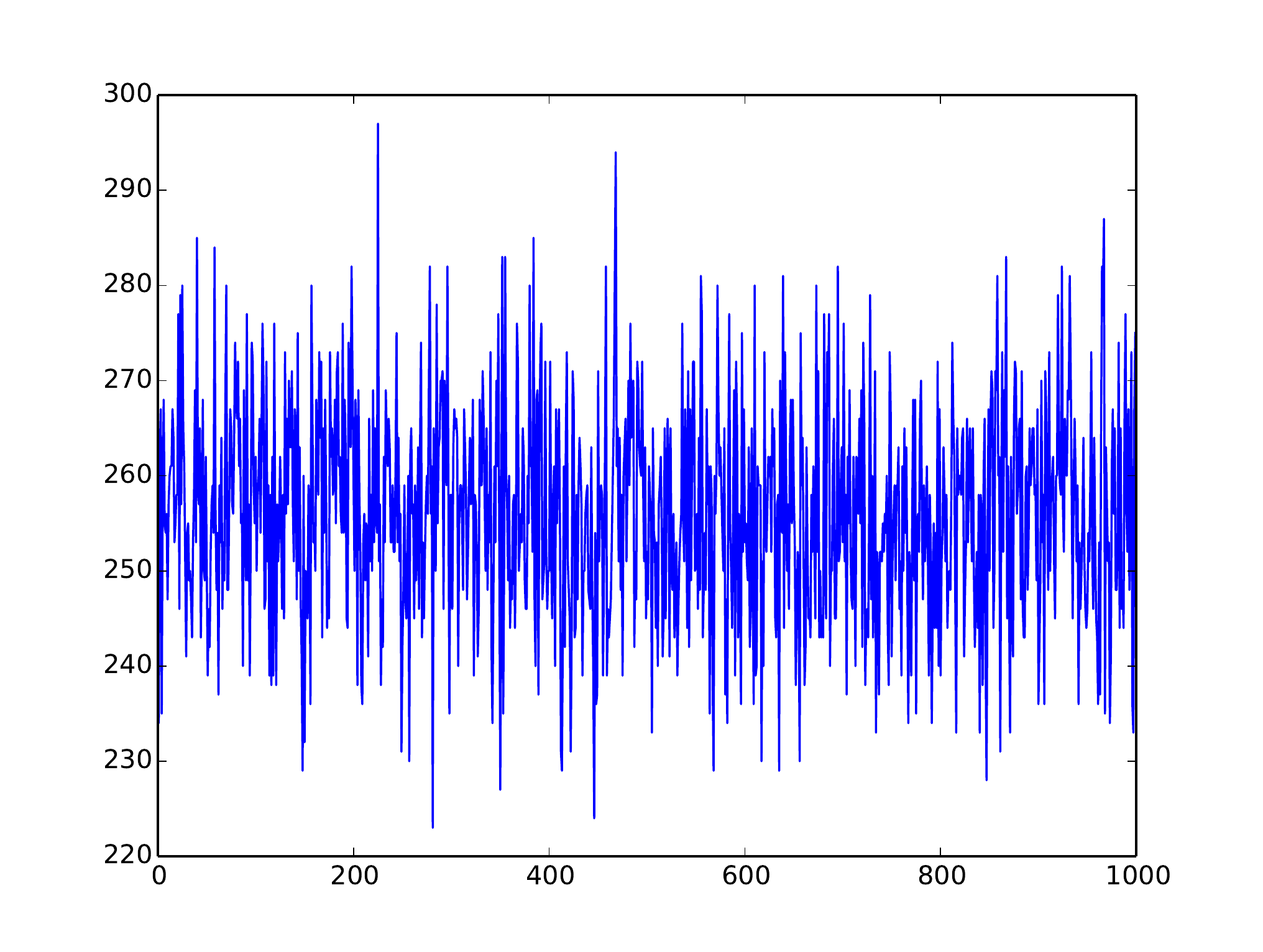}} \\
	\caption{Distribution of changed bit numbers $B_i$ with different standard hash functions}
	\label{fig3}
\end{figure}

Let us check the diffusion and confusion of the proposed hash function. The test procedure is described below. Firstly, we obtain the original hash value of plain text. Then at each time, only one bit is changed in it. The hash values of these modified plain texts are used to compare with the original hash value. After $N$ tests, $\overline{B}, P, \Delta B$, and $\Delta P$ are calculated. The key used here is $k_1=50,k_2=2,k_3=50, N=1000$. As shown in Figure \ref{fig3}, we check the distribution of changed bits of proposed hash function with MD5 and SHA-512. From these figures, we can see that one bit changed in the plain text can cause about $L/2$ modifications in $B_i$.

\begin{table}[htb]
	\caption{Statistical performance of the proposed hash function (B.B.S.)}
	\centering
	%\caption{Statical performance of the proposed hash function (B.B.S.)}
	\label{Tab1}
	{\begin{tabular}{cccccc}
			\hline
			$hash\_type$             & Iterate Times &$\overline{B}$     & $P(\%)$ & $\Delta B$ & $\Delta P(\%)$ \\ \hline
			\multirow{3}{*}{MD5}     & 1             & 63.906  & 49.926  & 5.819      & 4.546          \\
			& 2             & 63.845  & 49.879  & 5.656      & 4.419          \\
			& 10            & 63.846  & 49.880  & 5.845      & 4.566          \\ \hline
			\multirow{3}{*}{SHA-1}   & 1             & 79.774  & 49.859  & 6.446      & 4.029          \\
			& 2             & 80.355  & 50.222  & 6.329      & 3.956          \\
			& 10            & 79.779  & 49.862  & 6.131      & 3.832          \\ \hline
			\multirow{3}{*}{SHA-224} & 1             & 112.087 & 50.039  & 7.619      & 3.401          \\
			& 2             & 112.038 & 50.017  & 7.297      & 3.257          \\
			& 10            & 111.883 & 49.948  & 7.268      & 3.244          \\ \hline
			\multirow{3}{*}{SHA-256} & 1             & 128.075 & 50.029  & 7.845      & 3.064          \\
			& 2             & 127.72  & 49.891  & 8.002      & 3.126          \\
			& 10            & 127.806 & 49.924  & 8.215      & 3.209          \\ \hline
			\multirow{3}{*}{SHA-384} & 1             & 192.098 & 50.255  & 9.579      & 2.495          \\
			& 2             & 192.193 & 50.050  & 9.693      & 2.524          \\
			& 10            & 191.843 & 49.959  & 9.704      & 2.527          \\ \hline
			\multirow{3}{*}{SHA-512} & 1             & 256.043 & 50.008  & 10.867     & 2.122          \\
			& 2             & 256.062 & 50.012  & 11.376     & 2.222          \\
			& 10            & 256.032 & 50.006  & 11.438     & 2.234          \\ \hline
	\end{tabular}}
\end{table}

Observing Table \ref{Tab1}, even the iteration times are small, the mean changed bit numbers $\overline{B}$ and the mean changed probabilities $P$ are close to the desired values $L/2$ and $50\%$. $\Delta B$ and $\Delta P$ are quit small. In other words, the proposed hash function achieves desired values for such properties. These results illustrate the diffusion and confusion of the proposed hash function $H_h$, and these capabilities are quite stable. This feature is attributed to GFCIs that can change multi bits in one time. To sum up, due to the fact that computational complexity can be reduced here, we think it is better to apply it into practical applications.  Furthermore, compared with the performance of standard hash functions, which is shown in Table \ref{Tab2}, the proposed hash function shows better results.

\begin{table}[htb]
	\caption{Statistical performance of the standard hash function}
	\centering
	%\caption{Statical performance of the standard hash function}
	\label{Tab2}
	{\begin{tabular}{cccccccc}
			\hline
			$hash\_type$ & $\overline{B}$    & $P(\%)$ & $\Delta B$ & $\Delta P(\%)$ \\ \hline
			MD5          & 63.893  & 49.916  & 5.437      & 4.248          \\
			SHA-1        & 79.770  & 49.856  & 6.359      & 3.975          \\
			SHA-224      & 112.284 & 50.127  & 7.324      & 3.270          \\
			SHA-256      & 127.746 & 49.901  & 8.405      & 3.283          \\
			SHA-384      & 191.81  & 49.951  & 10.036     & 2.613          \\
			SHA-512      & 256.084 & 50.016  & 11.232     & 2.194          \\ \hline
	\end{tabular}}
\end{table}

\subsection{Collision analysis}

We now consider the analysis of impact resistance attacks. If hash function's ability to face collision is stronger, then the hash function is more security. Through experiments can be quantitatively tested the collision resistance ability of the proposed hash function. Firstly, we obtain the original hash value of plain text and transform it to ASCII code. Then the plain text one bit modification is applied and we obtain a new hash value in ASCII code. By comparing these two hash values, we can get the positions where they have the same character. The absolute coefficient between these two hash values can be described as follows:
\begin{equation}
d=\sum_{i=1}^N |t(e_i)-t(\check{e_i})|,
\end{equation}
where $N$ denotes the number of ASCII characters in the hash value, $e_i$ and $\check{e_i}$ are the $i^{th}$ character in former and new hash value separately. Function $t(\cdot)$ is used to transform $e_i$ and $\check{e_i}$ to decimal format. the theoretical value of average absolute distance per character is $85.3333$.

\begin{table}[htb]
	\caption{Collision performance}
	\centering
	\label{Tab4}
	{\begin{tabular}{ccccc}
			\hline
			$hash\_type$      & Number of hits  & sum of $d$ & avg $d$ per character   \\ \hline
			MD5               & (1931, 114, 3, 0, 0)  & 2956780    & 90.234                \\
			SHA-1              & (1880, 159, 8, 1, 0)  & 3472244    & 84.772                \\		
			SHA-224            & (1837, 204, 7, 0, 0)  & 4629075    & 80.725                \\			
			SHA-256            & (1817, 214, 17,0, 0)  & 5348270    & 81.608                \\			
			SHA-384            & (1690, 328, 27,3, 0)  & 8398092    & 85.430                \\			
			SHA-512           & (1601, 392, 47,6, 2)  & 11042728   & 84.250                \\   \hline		
	\end{tabular}}
\end{table}

The key used here is $k_1=50,k_2=2,k_3=50$, while testing times is $2048$. The experiment results are shown in Table \ref{Tab4}. The second column shows the number of hits, in which the fist component is the number of hits to zero, the second component is to one, the third component is to two, the forth one is to three, and the last one is to four. We can see that the maximum number of hits is four with small probability. It is mainly in the number of collision to zero and one. On the other hand, the average absolute difference $d$ of the two hash values per character, which is shown in the fifth column, is close to the desired value $85.3333$. Based on these results, the collision resistance capability of the proposed hash algorithm is strong.

\section{Conclusion}
In this article, a chaotic iteration based hash function has been presented. When constructing strategies, pseudorandom number generator is used. Then the general formulation of chaotic iterations is exploited to obtain hash values. Through the experimental evaluation of hash values, we can see that the proposed hash function is highly sensitive to initial parameters, initial values, and keys. The statistical performances show that the proposed hash function has better features of diffusion and confusion even if the iteration times are small, which can be considered for practical applications. And the proposed hash algorithm has better performance of collision performance. To sum up, the proposed scheme is believed a good application example for constructing secure keyed one-way hash function.

\bibliographystyle{splncs03}
\bibliography{Ref}
\end{document}